\documentclass[aps,prl,twocolumn,superscriptaddress,10pt,floatfix]{revtex4-2}
\usepackage[colorlinks=true,citecolor=blue,linkcolor=blue,urlcolor=blue,]{hyperref}
\usepackage{graphicx, nicefrac, textcomp}
\usepackage{float}
\usepackage[normalem]{ulem}
\usepackage{changes}
\usepackage{chngcntr}
\usepackage{amssymb}
\usepackage{soul}
\usepackage{color}
\usepackage{textcomp}
\usepackage{amstext}
\usepackage{graphicx}
\usepackage{gensymb}
\usepackage{graphics}\usepackage{subfigure}\usepackage{longtable}\usepackage{pstricks}\usepackage{dcolumn}\usepackage{bm}
\usepackage{siunitx}
\usepackage{booktabs}
\usepackage{multirow}
\usepackage{graphicx}
\usepackage{nicefrac}
\usepackage{xcolor}
\usepackage{rotating}
\begin{document}

\title{Floating zone growth at high oxygen pressures in Ruddlesden-Popper bilayer nickelate Y$_{y}$Sr$_{3-y}$Ni$_{2-x}$Al$_{x}$O$_{7-\delta}$}

\author{H.~Yilmaz}
\email[]{hasan.yilmaz@imw.uni-stuttgart.de}
\affiliation{University of Stuttgart, Institute for Materials Science, Materials Synthesis Group, Heisenbergstra{\ss}e 3, 70569 Germany}
\author{P.~Sosa-Lizama}
\affiliation{Max Planck Institute for Solid State Research, Heisenbergstra{\ss}e 1, 70569 Stuttgart, Germany}
\author{M.~Knauft}
\affiliation{Max Planck Institute for Solid State Research, Heisenbergstra{\ss}e 1, 70569 Stuttgart, Germany}
\author{K.~K\"uster}
\affiliation{Max Planck Institute for Solid State Research, Heisenbergstra{\ss}e 1, 70569 Stuttgart, Germany}
\author{U.~Starke}
\affiliation{Max Planck Institute for Solid State Research, Heisenbergstra{\ss}e 1, 70569 Stuttgart, Germany}
\author{M.~Isobe}
\affiliation{Max Planck Institute for Solid State Research, Heisenbergstra{\ss}e 1, 70569 Stuttgart, Germany}
\author{O.~Clemens}
\affiliation{University of Stuttgart, Institute for Materials Science, Materials Synthesis Group, Heisenbergstra{\ss}e 3, 70569 Germany}
\author{P. A.~van Aken}
\affiliation{Max Planck Institute for Solid State Research, Heisenbergstra{\ss}e 1, 70569 Stuttgart, Germany}
\author{Y. E.~Suyolcu}
\affiliation{Max Planck Institute for Solid State Research, Heisenbergstra{\ss}e 1, 70569 Stuttgart, Germany}
\author{P.~Puphal}
\email[]{puphal@fkf.mpg.de}
\affiliation{Max Planck Institute for Solid State Research, Heisenbergstra{\ss}e 1, 70569 Stuttgart, Germany}
\affiliation{2nd Physics Institute, University of Stuttgart, 70569 Stuttgart, Germany}

\date{\today}

\begin{abstract}
With the discovery of superconductivity under pressure in the Ruddlesden-Popper (RP) bilayer La$_3$Ni$^{2.5+}_2$O$_7$ and trilayer La$_4$Ni$^{2.66+}_3$O$_{10}$, a new field of nickelate superconductors opened up. In this respect, Sr-Ni-O RP-type phases represent alternative systems that exist with partial cation substitution. We demonstrate that by Y-doping in Sr$_{3}$Ni$_{2-x}$Al$_{x}$O$_7$ (SNAO), as Y$_{y}$Sr$_{3-y}$Ni$_{2-x}$Al$_{x}$O$_7$ (YSNAO), the drawback of an insulating ground state is overcome, and a significant decrease in resistivity is achieved with crystals exhibiting semiconducting behavior. We employ optical floating zone (OFZ) growth at 10 bar oxygen partial pressure to explore the phase formation in a narrow region of Y-Sr-Ni-Al-O and investigate via DFT the general stability of the pure Sr-Ni-O scenario. Using extensive diffraction and spectroscopy, as well as transport and magnetization measurements, the structural, chemical, electrical, and magnetic properties of the as-grown and reduced compounds were investigated. The optimal growth of YSNAO allows for large high quality crystals suitable for neutron studies. In the Al-free growth, a known $n=1$ RP system with Sr$_{1.66}$Y$_{0.33}$NiO$_{4-\delta}$, from which the first single crystals were obtained, was further confirmed, opening the door for future exploration of \textit{A}-site substituted RP-type phases without Ni-site disorder.
\end{abstract} 

\maketitle

\section{Introduction}

Nickelates provide an emerging platform for oxide superconductors with various structural \textit{motifs}: moving from hole-doped  rare-earth ($RE$) infinite-layer structures in $RE$NiO$_2$ \cite{Li2019}, to reduced RP Nd$_6$Ni$_5$O$_{12}$ \cite{Pan2021}, to various undoped RP-type phases under pressure. La$_3$Ni$_2$O$_7$ was found to show filamentary superconductivity both in the bilayer \cite{Sun2023} and in the new structural polymorph monolayer-trilayer variant \cite{Puphal2024}. Later on, also La$_4$Ni$_3$O$_{10}$ \cite{Zhu2024} and La$_2$PrNi$_2$O$_7$ \cite{Wang2024} were shown to bulk superconduct. Finally, even another polymorph of monolayer-bilayer shows supercondcutivity with yet a different T$_C$ \cite{Shi2025b}. In these RP-type phases, the superconductivity occurs at similar pressures of around 15~GPa, accompanied by a structural transition from an orthorhombic to a tetragonal system. The RP series denotes structural building blocks in oxide systems consisting of alternating rock-salt and perovskite layers with the composition $A_{n+1}T_{n}\mathrm{O}_{3n+1}$, where $n$ is the number of consecutive $T\mathrm{O}_{6}$ perovskite blocks within a structural unit. Tetragonal $I/4mmm$ RP-type phases are known for Sr on the $A$-site with different $3d$ or $4d$ transition metal ions \cite{Mitchell1998,CastilloMartinez2007,Sharma2004,Itoh1991,Dann1992,Elcombe1991,Mueller‐Buschbaum1990,Li2006,Li2020,Fukasawa2023}. With $A$ = Sr, the series implies the presence of a $4+$ oxidation state for the $T$-cations under a stoichiometric oxygen content. Due to the flexibility of the oxidation state of transition metals, a reduction down to Sr$_3T_2$O$_5$ is possible for certain compounds \cite{Schmidt2001}. Similarly, the $A$-site can be substituted by rare-earths \cite{Sharma1995}, leading to a lower oxidation state of the $T$ cation. However, only for $T=\mathrm{Ni}$ a full $A=$ La phase is reported for the whole RP-type phases \cite{Zhang1994}, leading to different oxidation states varying with $n$. In contrast, the Sr-bearing series has been much less studied, leaving room for its exploration and the possibility of stabilizing an ambient pressure superconductor.

The OFZ technique is one of the major ways in obtaining these new nickelate superconductors. Conventionally, phase diagrams are calculated or explored via differential thermal analysis (DTA) studies and subsequently OFZ growth is executed \cite{Wolff2022}. The technique as such has been previously used to explore phase diagrams and phase formations directly \cite{Shindo1980}. With more than quasy binary phases, e.g. in cuprate superconductors it can prove useful \cite{Kuroda1997,Kimura1992}. With the exploration of perovskite nickelates and solubility limits in doping \cite{Puphal2023APL}. Due to the slow process allowing for self seeding and zone purification, even impurity phases might crystallize large enough to allow for structure determination.

Here, we employ OFZ and analyze the obtained phases for Y-doping in the $A$-sites, i.e., $\mathrm{Y}_{y}\mathrm{Sr}_{3-y}\mathrm{Ni}_{2-x} \mathrm{Al}_{x}\mathrm{O}_7$. A unique crystal formation exhibiting a tetragonal structure without external pressure is examined, which is structurally analogous to recently established RP-type $n=2$ single crystals for Sr-Ni-O with Al-substitution (SNAO) \cite{Yilmaz2024}. Unlike Sr$_{3}$Ni$_{2-x}$Al$_{x}$O$_7$, with Y-doping transitions occurred solely between stable RP phases with varying $n$. Notably, by leveraging $\mathrm{Y}^{3+}$ doping, conductivity is drastically increased by a factor of $10^6$. While X-ray diffraction (XRD) and scanning transmission electron microscopy (STEM) measurements prove the structural quality of these crystals, X-ray photoemission spectroscopy (XPS) experiments combined with thermogravimetric analysis (TGA) reveal an extraordinarily high oxidation state. Additionally, upon topochemical oxygen reduction, depending on the doping content, reduced phases of varying oxidation states can be reached.

\section{Methods}
The experimentally obtained crystal structure was analyzed using Density functional theory (DFT). DFT calculations were performed using the scalar-relativistic implementation of the full-potential local orbital (FPLO) code \cite{lit:Koepernik1999_FPLO}, version 22.00-62. Self-consistency was achieved when the change in total energy between subsequent steps was less than $10^{-8}$\,Hartee. The exchange-correlation potential was parametrized as described by Perdew and Wang \cite{lit:Perdew1992_Perdew_Wang_XC} and the Brillouin zone integration carried out on regular meshes chosen such that the total energy was converged to better than 1\,meV.

The phase diagram was constructed by calculating the formation energy based on references of the elemental solids of Sr, Ni and $\mathrm{O_2}$. Experimental crystal structures were  taken from the Inorganic Crystal Structure Database (ICSD) \cite{lit:Zagorac_ICSD, lit:phaseDiagram_Ni,lit:phaseDiagram_NiO,lit:phaseDiagram_Sr,lit:phaseDiagram_SrO,lit:phaseDiagram_SrO2,lit:phaseDiagram_O2,lit:phaseDiagram_SrNiO2}. For the case of $\mathrm{Sr_2NiO_4}$ and $\mathrm{Sr_3Ni_2O_7}$, we started from the experimentally determined structures (Ref. \cite{James1995} and present work) with doping and manually removed dopants to simulate the parent compound. For $\mathrm{Sr_9Ni_7O_{21}}$, the partially occupied oxygen ions were moved to the high symmetry site of the structure reported in Ref. \cite{lit:phaseDiagram_Sr9Ni7O21}. The algorithm for the convex hull is implemented in the atomic simulation environment (ASE) \cite{HjorthLarsen2017}.

The aim of synthesis is to grow differently Y doped Sr$_3$Ni$_2$O$_7$ crystals, where co-doping of Al is necessary \cite{Yilmaz2024}, which is achieved by preparing different precursor phases and performing floating zone growth. For the aimed Y$_{y}$Sr$_{3-y}$Ni$_{2-x}$Al$_{x}$O$_7$ crystals, the optical floating zone (OFZ) technique was used under 9~bar oxygen partial pressure. First, feed rods were prepared by the conventional Solid-State method using stoichiometric amounts of $\mathrm{Y}_{2}\mathrm{O}_3$ (Alfa Aesar, 99.995\%), $\mathrm{SrCO}_3$ (Sigma Aldrich, 99.995\%), $\mathrm{NiO}$ (Alfa Aesar, 99.998\%), and $\mathrm{Al}_{2}\mathrm{O}_3$ (Alfa Aesar, 99.995\%). The precursor powders were mixed using a ball mill and sintered in an aluminum oxide crucible at 1100~\textdegree C for 12~hours in air. The as-sintered powders were formed into cylindrical rods (\(\phi\)=4~mm; $l$=100~mm) by an isostatic press (70~MPa) using rubber forms and subsequently annealed at 1200~\textdegree C 12~hours in air. The rod was installed in a classical horizontal four mirror lamp Optical Float Zone Furnace (FZ-T-10000-H-III-VPR). Finally, single crystal growth was conducted with 4$\times$1000~W halogen lamps. During the growth process, the feed and seed rods were counter-rotated (24~rpm) in order to minimize the diffusion zone close to the crystallization surface. Oxygen gas flow (100~cc/min) was applied, and an auxiliary pressure of 9~bar was simultaneously introduced to stabilize the highly oxidized phase. To obtain relatively large single crystals, the growth rate was kept at 4~mm/h. After the growth, the boule was cracked using a mortar, and single crystals were selected for further analysis. 

\begin{figure*}[t]
\centering
\includegraphics[width=1.7\columnwidth]{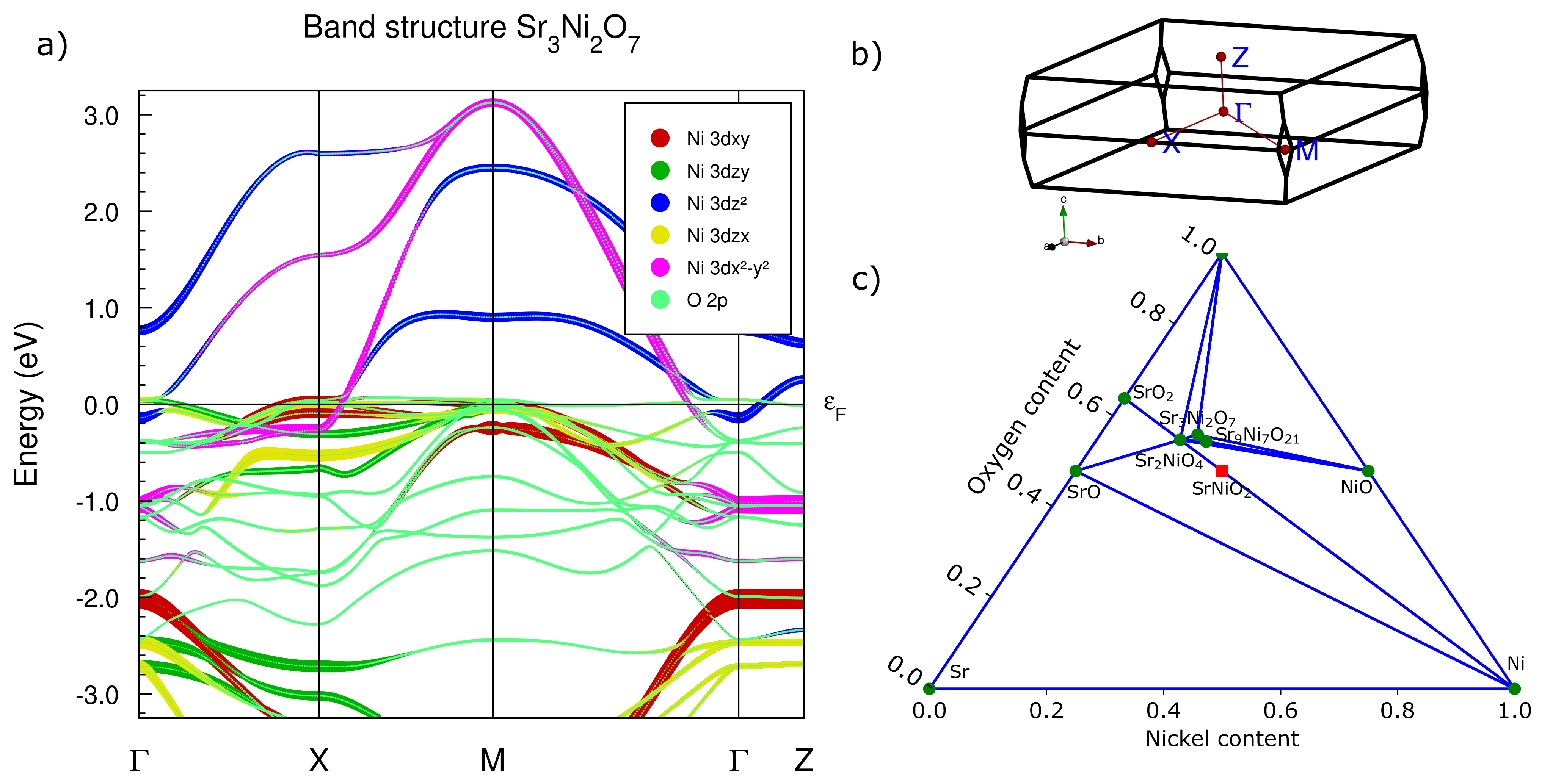}
\caption{(a) Orbital-projected band structure for Ni $3d$- and O $2p$-orbitals calculated by DFT. (b) High-symmetry path in the Brillouin zone of $\mathrm{Sr_3Ni_2O_7}$. (c) Ternary phase diagram of Sr-Ni-O$_2$ based on stability calculations. Stable (unstable) phases are marked as green (red) points.
}
\label{DFT}
\end{figure*}

The crystallinity and phase purity of the as-grown single crystals were examined by XRD on powdered samples. Powder X-ray diffraction (PXRD) measurements were performed on ground crystals in order to investigate the presence of impurities as well as to determine the crystal structure and lattice parameters. XRD patterns were recorded at room temperature using a Rigaku Miniflex diffractometer with Bragg-Brentano geometry, Cu~$K_{\alpha}$ radiation, and a Ni filter. Rietveld refinements were carried out with TOPAS V6 software. Single crystal diffraction was performed at room temperature using a Rigaku XtaLAB mini II instrument with Mo~$K{_\alpha}$ radiation. Data were analyzed with CrysAlis (Pro), and the final refinement was performed using Olex2 with SHELX.  The x-ray Laue diffraction images were collected with a Photonic Science CCD detector using a standard W broad X-ray source operated at 35~kV and 40~mA. Energy-dispersive X-ray spectra (EDX) were recorded with a NORAN System 7 (NSS212E) detector in a Tescan Vega SEM (TS-5130MM) and crosscheck with an additional measurement in a JSM-IT210. On each crystal, dozens of points were analyzed, and the standard deviation was considered as an error.  The STEM lamellae were prepared by thinning the crystal specimen down to electron transparency via Focused Ion Beam milling with Ga ions on a Thermo Fisher Scios II FIB using the liftout method. The lateral dimensions of the specimen were $12 \times 10 ~\mu\mathrm{m}$, with an overall thickness below 100~nm. In the analyzed regions of the samples, the thickness was $\sim 50$~nm determined by the Fourier log-ratio method. High-resolution STEM imaging was performed with a JEOL JEM-ARM200F equipped with a cold field-emission electron source and a probe $C_\mathrm{s}$-corrector (DCOR, CEOS GmbH), at an acceleration voltage of 200 kV. The STEM-HAADF acquisition was carried out with a convergence (semi-) angle of 28~mrad, resulting in a probe size of $\sim 0.8 \textup{~\AA}$. Collection angles for HAADF images were 75 to 310~mrad. 
XPS measurements were carried out at a Kratos Axis Ultra with a monochromated Al~$K_\alpha$ X-ray source. The small crystals were glued on a dedicated sample holder with pilars (with a diameter smaller than the single crystals) in order to avoid signals from the environment. The crystals were cleaved in the glove box revealing clean shiny ab-plane surfaces. The samples were transferred under inert gas into the XPS system without exposure to air.  A charge neutralizer was used to compensate for surface charging, and the binding energy was calibrated to C~$1s$ at 284.8~eV (adventitious carbon). The data was analyzed using Casa XPS ~V2.3.26rev1 \cite{Fairley2021} and the peaks were fitted with a LA (Gaussian and Lorentzian convolution) line shape after Shirley background subtraction for the quantification.
Differential Thermal Analysis (DTA) and Thermo gravimetry Analysis (TGA) were performed using a Netzsch DTA/TG.
Magnetic susceptibility measurements were conducted using a vibrating sample magnetometer (MPMS~3, Quantum Design) on a 257~mg  $x=0.25$ and a 11~mg  $x=0.166$ YSNAO oriented platelike single crystal. Electrical transport measurements were carried out using a Physical Property Measurement System (PPMS, Quantum Design). The samples were 3 x 4 x 0.3~mm$^3$ $ab-$plates with a four point contact glued by silver paint in a line oriented along the 110 direction, which is the cleaved long direction of the platelike crystals.

\section{Results}
\subsection{Electronic structure and phase stability}

Motivated by our previous synthesis of $n = 2$  Sr-based RP nickelates via successful Al substitution in SNAO \cite{Yilmaz2024}, simple DFT calculations for the Al free RP $n=2$ Sr$_3$Ni$_2$O$_7$ case were performed based on experimental structure results with Al crystallizing in the tetragonal $I4/mmm$ structure. 
The results are displayed in Fig.~\ref{DFT}~(a) with individual d orbitals colorcoded. Notably, the overall bandstructure is (not surprisingly) extremely similar to bilayer La$_3$Ni$_2$O$_7$ under external pressures of  29.5~GPa \cite{Cao2024} besides an enormous shift of the Fermi energy of 1~eV down due to the overall strong doping realized with Sr$^{2+}$ on the $A$-site. Here, however without external pressure application a tetragonal structure is realized and with successful electron doping a similar superconductor could be created. DFT results indicate a metal with multiple bands crossing the Fermi level. These bands are primarily from Ni-$d_z^2$ and Ni-$d_{x^2-y^2}$ orbitals. In addition, various flat bands of Ni-$d_{xy}$ and Ni-$d_{xz}$ exist at the Fermi level. Notably, flat band systems are currently under intense investigation in quantum materials as they usually exhibit topological properties and their reduced bandwidth proportionally enhances the effect of Coulomb interaction \cite{Checkelsky2024,Rosa2024,Neves2024}. As such the strongly correlated Sr$_3$Ni$_2$O$_7$ system presents an interesting candidate. This DFT picture however is insufficient due to the strong electronic correlations of Ni-$d$ electrons, but gives a first insight into this new candidate of nickelate superconductor. The latter could be realized in future via doping pure Sr$_3$Ni$_2$O$_7$ with hope at ambient pressure superconductivity from the already tetragonal structure . Reaching ambient pressure superconductivity might be experimentally easier to achieve for bulk samples of Sr than for the La system, as the latter realizes for full oxidization an orthorhombic distorted structure \cite{Puphal2024}. Substituting La in order to dope the system would result in chemical pressure \cite{Lin2022}, which, in contrast to external pressure would increase this distortion. Here, one should mention recent thin film results, which can via epitaxy influence the bond angle and thus superconductivity at ambient conditions is achievable \cite{Ko2024}. Similarly, one could attempt to go into the Sr based nickelate regime via heavy Sr substitution, resulting in a closely related band structure with a 0.4~eV shift of the fermi energy at a stoichiometry of La$_2$SrNi$_2$O$_7$ (corresponding roughly to the expected 1/3 of our pure Sr calculations) \cite{Shi2025}. This doping however is quite limited in perovskites \cite{Puphal2023APL} by a solubility limit to 8~\% and hence also unlikely to be experimentally feasible for RP $n=2$.

Furthermore, stability calculations have been performed for the Sr-Ni-O phase diagram as shown in Fig.~\ref{DFT}~(c). While the binary phase diagram \cite{Zinkevich2004,Yilmaz2024} only reveal Sr$_9$Ni$_7$O$_{21}$, which we also found as the stable phase for a "Sr$_3$Ni$_2$O$_7$" OFZ growth at 10~bar O$_2$, calculations indicate a general stability of the simple RP $n=2$ Sr$_3$Ni$_2$O$_7$ and RP $n=1$ Sr$_2$NiO$_4$. Notably, this does not yield the experimental conditions to realize this phase and as showed before, Al substitution stabilizes the RP $n=2$ phase \cite{Yilmaz2024}. Interesting is also the stability of $n=1$ in the course of this work as it could be stabilized by simple $A-$site substitution, without the disadvantage of co-doping the Ni-site.

\subsection{Phase stability exploration}

\begin{figure}[h]
\centering
\includegraphics[width=1.0\columnwidth]{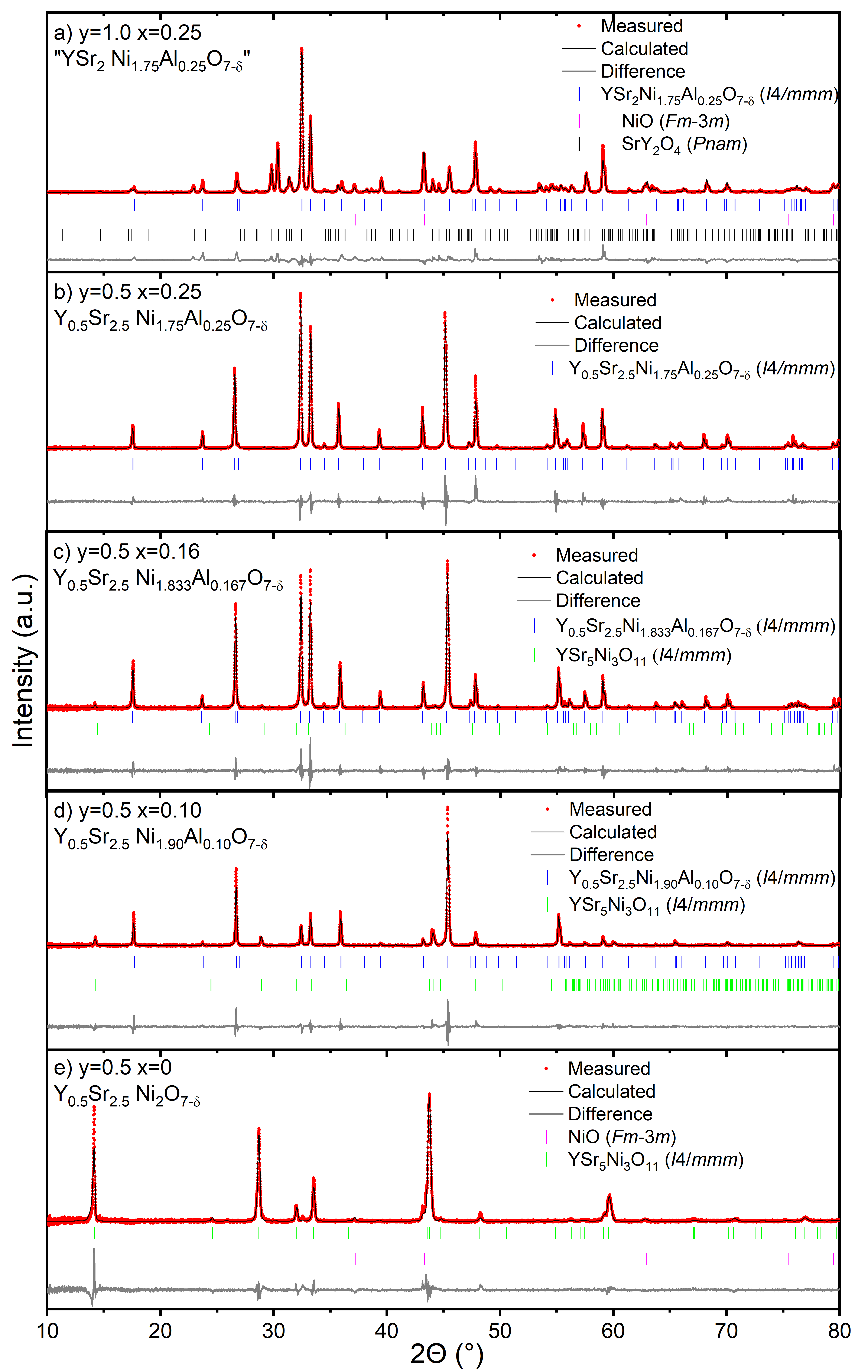}
\caption{Rietveld fit of the room-temperature PXRD pattern of $\mathrm{Y}_{y}\mathrm{Sr}_{3-y}\mathrm{Ni}_{2-x}\mathrm{Al}_{x}\mathrm{O}_{7-\delta}$. For each plot, the solid black line corresponds to the calculated intensity from the Rietveld refinement, the solid gray line is the difference between the experimental and calculated intensities, and the vertical blue/green bars are the calculated Bragg peak positions.
}
\label{XRD}
\end{figure}

In this work, the effect of electron doping was extensively explored by introducing Y into the structure as $\mathrm{Y}_{y}\mathrm{Sr}_{3-y}\mathrm{Ni}_{2-x}\mathrm{Al}_{x}\mathrm{O}_{7-\delta}$ (YSNAO). The crystals are grown after the preparation of stoichiometric powders, as described in the methods section. First, a greater doping with $y = 1$ was attempted with the established content of substitutional Al, $x = 0.25$. Here, the growth required large feeding rates but otherwise realized a stable optical floating zone process. However, the obtained boule revealed a phase mix, as visible via the PXRD results shown in Fig.~\ref{XRD}~(a), and contained only small crystals of sub-millimeter size. Most importantly, the $n = 2$ RP-type phase formed as the main phase, with reduced lattice constants (see Table~\ref{tab:lattice_comp}) as expected from the smaller Y ions (see Table~\ref{tab:lattice_comp}). Additionally, the Y solubility limit was reached as visible from the presence of a Y-rich impurity phase of the spinel phase SrY$_2$O$_4$ (see Fig.~\ref{XRD}~(a) and Table \ref{tab:lattice_comp}) with 22.98(14)~wt\%. As the boule contained shiny single crystals regardless of the presence of impurity phases, the stoichiometry of the RP phase could be analyzed by EDX and found to be Y$_{0.47(5)}$Sr$_{2.56(3)}$Ni$_{1.78(8)}$Al$_{0.27(2)}$O$_{6.5(7)}$. 

\begin{figure}[h]
\centering
\includegraphics[width=1.0\columnwidth]{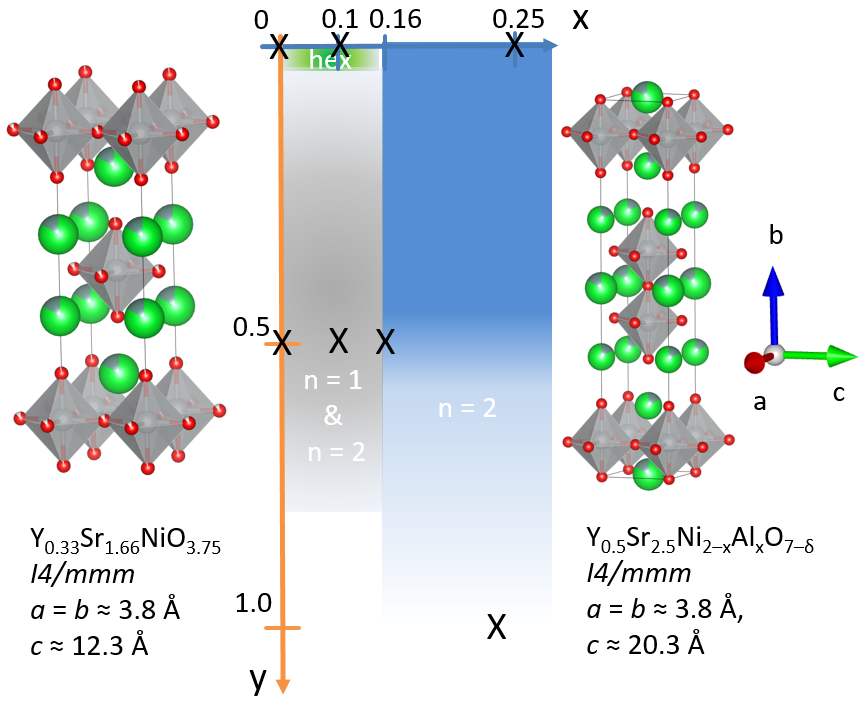}
\caption{Chemical stability of the Y$_{y}$Sr$_{3-y}$Ni$_{2-x}$Al$_{x}$O$_7$ series, with the  Al (Y) content $x$ ($y$) as the horizontal (vertical) axis. At the axis, i.e. with finite $y$ at $x = 0$, the defect Ruddlesden-Popper $n = 1$ type-phase Y$_{0.33}$Sr$_{1.66}$NiO$_{3.75}$ (or YSr$_5$Ni$_3$O$_{11}$) and NiO are formed. For $0 <  x < 0.166$ and $0.5 < y \leq 1$ (grey region), a mixture of the two phases is found. Notably, the narrow green stripe marks the Y-free area, where Sr$_9$Ni$_7$O$_{21}$ is the phase mixing with $n=2$ \cite{Yilmaz2024}, while in the blue range, phase pure $n = 2$ RP-type Y$_{0.5}$Sr$_{2.5}$Ni$_{2-x}$Al$_{x}$O$_7$ is obtained (see Figure \ref{XRD}).
}
\label{view}
\end{figure}

\begin{table*}
\caption{Result of Rietveld powder refinement. The growth are sorted by their success, where a solubility limit is denoted by X and a successful growth with sizeable single crystals by \checkmark. Listed are the attempted or nominal composition, with their substitution content for Y ($y$) and Sr ($x$),  the relative wt\% of the RP-type phase $n$, the EDX results of stoichiometry, the TGA result of oxygen content, the lattice parameters $a,b,c$ and the Nickel oxidation state from refinement, TGA and EDX results for different float zone growths and hydrogen-treated phases. Notably both RP $n=1,2$ crystallize in the tetragonal $I4/mmm$ (\#139) structure, but upon reduction become orthorhombic $Immm$ denoted by explanation marks around the RP $n$.}
\label{tab:lattice_comp}

\begin{tabular}{@{}l@{}l@{}l@{}l@{}l@{}l@{}lccccc}
 & \multicolumn{2}{c}{\textbf{\scriptsize{}{Attempt}}} & \textbf{\scriptsize{}{}wt\%}{\scriptsize{}{} }  & \textbf{\scriptsize{}{}$n$}{\scriptsize{} } & \textbf{\scriptsize{}{}EDX}  & \textbf{\scriptsize{}{}TGA}  & \textbf{\scriptsize{}{}a{} {[}\AA{]} }{\scriptsize{} } & \textbf{\scriptsize{}{}b{} {[}\AA{]} }{\scriptsize{} } & \textbf{\scriptsize{}{}c{} {[}\AA{]}  } & \multicolumn{2}{c}{\textbf{\scriptsize{}{}Ni$^{x+}$}}\tabularnewline
 & {\scriptsize{}{}$y$}  & {\scriptsize{}{}$x$}  &  &  &  & {\scriptsize{}{}O$_{x}$}  &  &  &  & {\scriptsize{}{}EDX}  & {\scriptsize{}{}TGA} \tabularnewline
\midrule 
\multirow{4}{*}{X~~} & \textbf{\scriptsize{}{}1}{\scriptsize{} } & \textbf{\scriptsize{}{}0.25~}{\scriptsize{} } & {\scriptsize{}{}72.38(18) }  & {\scriptsize{}{}2 }  & {\scriptsize{}{}Y$_{0.47(5)}$Sr$_{2.56(3)}$Ni$_{1.78(8)}$Al$_{0.27(2)}$O$_{6.5(7)}$}  &  & {\scriptsize{}{}3.79089(6) }  & {\scriptsize{}{}-}  & {\scriptsize{}{}19.8549(4) }  & {\scriptsize{}{}3.2}  & \tabularnewline
\cmidrule{2-12} 
 & \textbf{\scriptsize{}{}0.5~~ }{\scriptsize{} } & \textbf{\scriptsize{}{}0 }{\scriptsize{} } & {\scriptsize{}{}77.5(11) }  & {\scriptsize{}{}1 }  & {\scriptsize{}{} Y$_{0.34(7)}$Sr$_{1.6(1)}$Ni$_{1.0(2)}$O$_{3.0(3)}$}  &  & {\scriptsize{}{}3.76333(15) }  & {\scriptsize{}{}-}  & {\scriptsize{}{}12.3791(3) }  & {\scriptsize{}{} 2.1 } & \tabularnewline
\cmidrule{2-12} 
 & \textbf{\scriptsize{}{}0.5}{\scriptsize{} } & \textbf{\scriptsize{}{}0.1 }{\scriptsize{} } & {\scriptsize{}{}90.0(4) }  & {\scriptsize{}{}2 }  & {\scriptsize{}{} Y$_{0.48(4)}$Sr$_{2.6(2)}$Ni$_{1.8(2)}$Al$_{0.13(1)}$O$_{6.9(4)}$}  &  & {\scriptsize{}{}3.79112(5) }  & {\scriptsize{}{}-}  & {\scriptsize{}{}19.9151(2) }  & {\scriptsize{}{} 3.8 } & \tabularnewline
 &  &  & {\scriptsize{}{}10.0(4) }  & {\scriptsize{}{}1 }  &  &  & {\scriptsize{}{}3.7887(17) }  & {\scriptsize{}{}-}  & {\scriptsize{}{}12.2871(3) }  &  & \tabularnewline
\midrule 
\multirow{5}{*}{\checkmark~~ } & \textbf{\scriptsize{}{}0.5}{\scriptsize{} } & \textbf{\scriptsize{}{}0.166~~}{\scriptsize{} } & {\scriptsize{}{}98.45(11) }  & {\scriptsize{}{}2 }  & {\scriptsize{}{}Y$_{0.55(1)}$Sr$_{2.52(2)}$Ni$_{1.83(6)}$Al$_{0.16(8)}$O$_{6.6(4)~~}$}  & {\scriptsize{}{}6.23}  & {\scriptsize{}{}3.79113(4) }  & {\scriptsize{}{}-}  & {\scriptsize{}{}19.92860(18) }  & {\scriptsize{}{} 3.3 } & {\scriptsize{}{}2.7} \tabularnewline
 &  &  & {\scriptsize{}{}1.55(11) }  & {\scriptsize{}{}1 }  &  &  & {\scriptsize{}{}3.8225(17) }  & {\scriptsize{}{}-}  & {\scriptsize{}{}12.2399(17) }  &  & \tabularnewline
 &  &  & {\scriptsize{}``100''}  & {\scriptsize{}``2'' }  &  & {\scriptsize{}{}5.29}  & {\scriptsize{}{}3.51772(5) }  & {\scriptsize{}{}3.85059(7) }  & {\scriptsize{}{}20.2461(4) }  & {\scriptsize{}{}1.75}  & {\scriptsize{}{}1.6} \tabularnewline
\cmidrule{2-12} 
 & \textbf{\scriptsize{}{}0.5}{\scriptsize{} } & \textbf{\scriptsize{}{}0.25}{\scriptsize{} } & {\scriptsize{}{}100 }  & {\scriptsize{}{}2 }  & {\scriptsize{}{}Y$_{0.51(6)}$Sr$_{2.52(3)}$Ni$_{1.74(2)}$Al$_{0.26(4)}$O$_{6.7(2)}$}  & {\scriptsize{}{}6.13}  & {\scriptsize{}{}3.78842(3) }  & {\scriptsize{}{}-}  & {\scriptsize{}{}19.99846(16) }  & {\scriptsize{}{}3.5}  & {\scriptsize{}{}2.875} \tabularnewline
 &  &  & {\scriptsize{}``100''}  & {\scriptsize{}``2'' }  &  & {\scriptsize{}{}4.89}  & {\scriptsize{}{}3.5506(4) }  & {\scriptsize{}{}3.8490(4) }  & {\scriptsize{}{}20.254(2) }  & {\scriptsize{}{}2}  & {\scriptsize{}{}1.5} 
\end{tabular}

\end{table*}

The next step was a growth with $x = 0.25$ and $y = 0.5$ under the same conditions as described in the methods section. The process required only subtle feeding, and the melt was extremely stable. Upon tapping of the obtained boule, crystals of several mm in length broke apart. For this case, PXRD in Fig.~\ref{XRD}~(b) reveals a pure $n = 2$ RP-type phase, with the corresponding lattice constants summarized in Table~\ref{tab:lattice_comp}. This is further confirmed by the exact stoichiometric result of EDX, indicating  Y$_{0.51(6)}$Sr$_{2.52(3)}$Ni$_{1.74(2)}$Al$_{0.26(4)}$O$_{6.7(2)}$. Moreover, due to the homogeneous crystals obtained, it was motivated to both reduce the Al-disorder in the Ni lattice and enhance the electric conductivity by a growth of $y = 0.5$ and $x = 0.166$. PXRD of the grown crystals are shown in Fig.~\ref{XRD}~(c), at first glance revealing a quite pure $n = 2$ RP-type phase, though with slightly smaller single crystals than in the previous case. The EDX result on the obtained $n=2$ crystals revealed a matching stoichiometry of Y$_{0.55(1)}$Sr$_{2.52(2)}$Ni$_{1.83(6)}$Al$_{0.16(8)}$O$_{6.6(4)}$. An extremely minor impurity of the RP $n=1$ phase can be identified marking the solubility limit in this stoichiometry. The compound is still listed as a successful growth in Table~\ref{tab:lattice_comp} and marks the line in the phase diagram in Fig.~\ref{view}

In addition, after a successful growth with a reduced Al content, another growth was conducted without Al, maintaining the same Y content, $y = 0.5$ and $x = 0$. Similarly, the growth was stable with the required feeding but resulted in a phase mix. PXRD analysis in Fig.~\ref{XRD}~(e) revealed a major formation of the defect RP $n = 1$-type phase $\mathrm{Y}_{0.33}\mathrm{Sr}_{1.66}\mathrm{NiO}_{3.75}$ (or $\mathrm{YSr}_{5}\mathrm{Ni}_{3}\mathrm{O}_{11}$\cite{James1993,James1995}) with 73.0(9)~wt\%. Thus despite our calculation of general stability of Sr$_3$Ni$_2$O$_7$ (see Fig.~\ref{DFT}~(c) the phase remains unstable even with the addition of Y, which at least stabilizes the $n=1$ Sr RP phase. Here, one should however note that the growths are executed at limited oxygen pressures of 10~bar and future high pressure experiments might reveal to be more successful.
Finally, an intermediate Al-concentration was attempted with $y = 0.5$  and $x = 0.1$, still yielding a phase mix. This time it consists of $n = 1$ and $n = 2$, as visible in the PXRD in Fig.~\ref{XRD}~(d). The major phase here, however is still $n = 2$ with 90~wt\%, as summarized in Table~\ref{tab:lattice_comp}, with a slightly reduced $c$-axis in comparison to $x = 0.166$. Due to the inter-growth of two phases, the crystal size is further reduced, but the two phases of crystals can be clearly distinguished in single crystalline form. Hence, the phase diagram of YSNAO can be drawn as depicted in Fig.~\ref{view}, where the transitions occur only between the RP-type phases when Y is introduced to the system. In addition a summary of the results is given in Table~\ref{tab:lattice_comp}, sorted by solubility limited growths marked with an X and successful growth marked with a checkmark.

 \subsection{Crystallinity, orientation and oxidation state}

\begin{figure}[h]
\centering
\includegraphics[width=1.0\columnwidth]{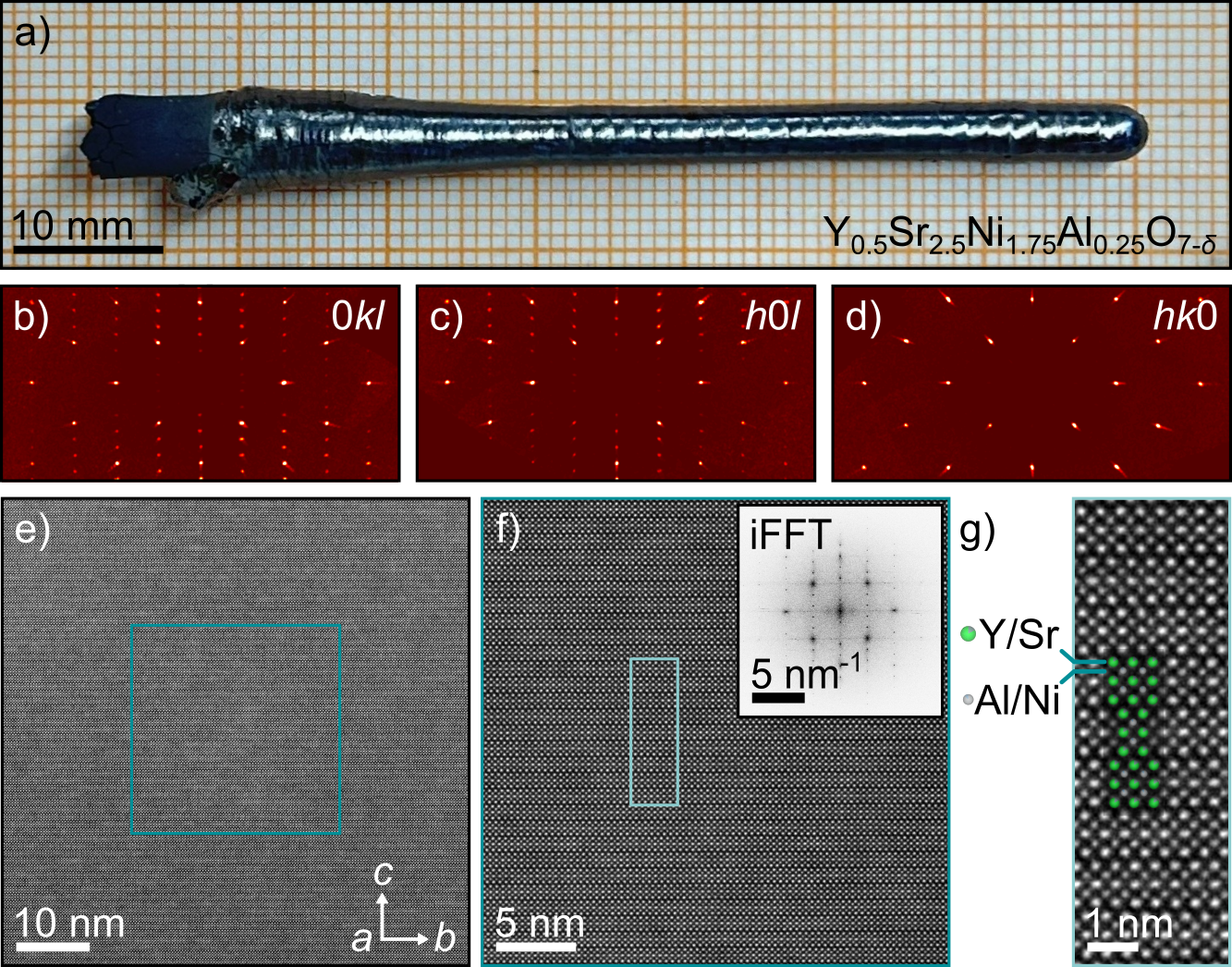}
\caption{(a) Picture of the OFZ grown boule of $\mathrm{Y}_{0.5}\mathrm{Sr}_{2.5}\mathrm{Ni}_{1.75}\mathrm{Al}_{0.25}\mathrm{O}_{7-\delta}$, and its single crystal XRD zonal maps along (b) $0kl$, (c) $h0l$ and (d) $hk0$. (e) Overview STEM-HAADF image of a region of $64\times64$ nm from the same crystal shows a defect-free sample. (f) High-magnification image showing the presence of a single stacking order, consisting of bilayer perovskite slabs. The inset corresponds to the iFFT of the image in (e), analogous to (b). (g) Enlarged region within (f) shows the clear RP-type $n=2$ stacking, where the larger (smaller) intensity spots correspond to atomic columns of Sr/Y (Ni/Al). The structural model is overlaid for reference.}
\label{STEM}
\end{figure}

For the phase pure Y$_{0.5}$Sr$_{2.5}$Ni$_{1.75}$Al$_{0.25}$O$_{7-\delta}$ crystal, the boule and single crystal images are shown in Fig.~\ref{STEM}~(a-d). The single-crystal diffraction refinement results corresponding to the zonal maps are summarized in Table~\ref{tab:SNOatom_pos}. Moreover, the structural quality of the crystal was verified on the nanoscale with scanning transmission electron microscopy (STEM) imaging. STEM high-angle annular dark field (HAADF) images at increasing magnifications show a clean, defect-free sample in Fig.~\ref{STEM}~(e-g). 
The higher-magnification image in Fig.~\ref{STEM}~(g) exemplifies the presence of the observed single RP-type phase (n = 2)  throughout the sample. Moreover, the inverted Fast Fourier Transform (iFFT) taken over the larger field-of-view shows the spatial frequencies consistent with the extinction rules for the $I4/mmm$ space group, which is in agreement with the $(0kl)$ XRD map (see Fig.~\ref{STEM}~(b)). Atomic columns of different sizes are observed when comparing $A$- and $T-$sites, with the former (Sr/Y) larger than the latter (Ni/Al), in accordance with the atomic number contrast of STEM-HAADF.

\begin{figure}[h]
\centering
\includegraphics[width=0.7\columnwidth]{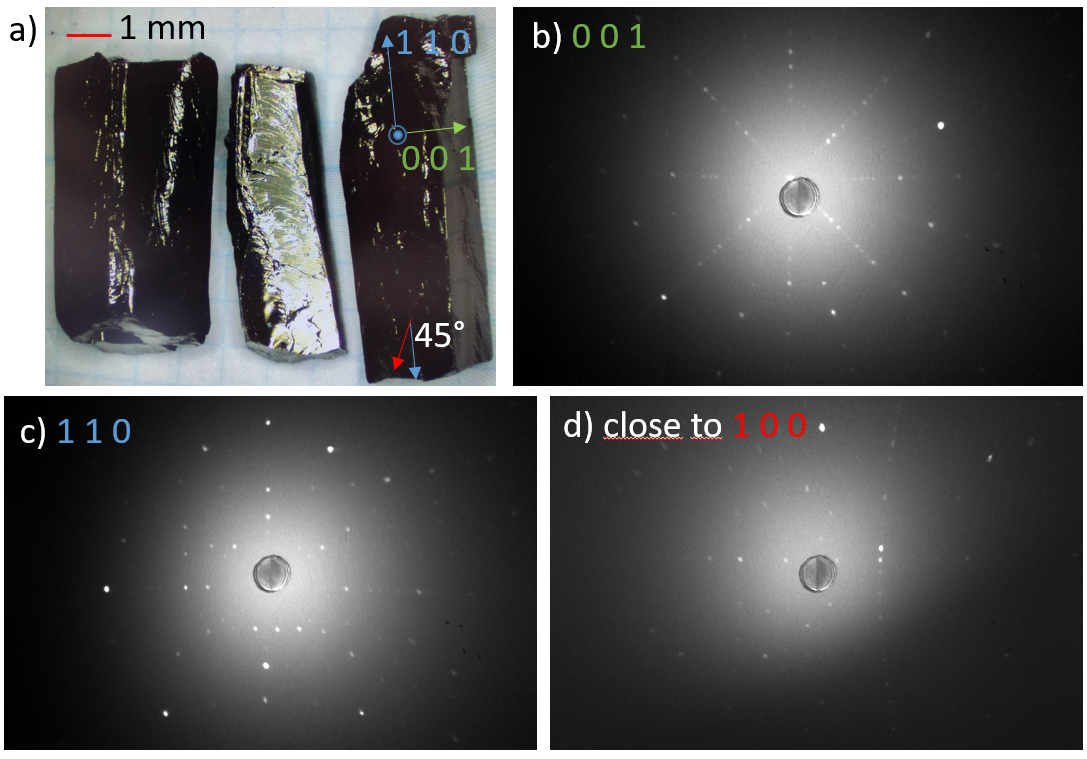}
\caption{Orientation and size of the obtained single crystals. (a) Image of broken single crystals with the typical orientation indexed on one. The corresponding Laue XRD images of cleaved (b) 001 direction (c) 110 direction and (d) the high symmetry but not visible 100 direction.
}
\label{Laue}
\end{figure}

By Laue diffradction the orientation and single crystallinity of the as grown boule and pieces separated by breaking were checked. The crystals grow along the 110 direction and break into elongated shiny plates that reveal the 001 direction usually as the largest surface. The in-plane cleaved orientations are the 110 directions, with the $a-$ and $b-$axes typically on the edges. Corresponding crystals are displayed in Fig.~\ref{Laue}~(a) and the Laue diffraction images of 001 in (b), 110 in (c) and the edge 100 direction in (d). The backsides of the crystals reveal perfectly matching Laue spots proving several mm sized single crystals.

\begin{table}[tb]
\caption{Refined atomic coordinates of (a) Y$_{0.5}$Sr$_{2.5}$Ni$_{1.75}$Al$_{0.25}$O$_{7-\delta}$ and (b) Y$_{0.5}$Sr$_{2.5}$Ni$_{1.833}$Al$_{0.166}$O$_{7-\delta}$ in the tetragonal $I4/mmm$ (\#139) structure, extracted from single crystal XRD. The corresponding lattice constants and quality parameters are: \newline
(a) $a=b=3.8045(4)$ {\AA}, $c=19.964(3)$ {\AA}; R$_1=0.0459$, w$_{R2}=0.1349$ and \newline
(b) $a=b=3.8147(2)$ {\AA}, $c=20.0197(19)$ {\AA}; R$_{1}=0.0263$, w$_{R1}=0.0843$ }
\label{tab:SNOatom_pos}
\begin{tabular}{cccccccccccc}

\textbf{a)} &   & & &   &  \\ \hline

\textbf{Atom} & \textbf{Site}  & \textbf{x} & \textbf{y} &  \textbf{z} & \textbf{Occ} \\ \hline
Y             & Y01            & $\nicefrac{1}{2}$        & $\nicefrac{1}{2}$        & 0.18219(9) & $\nicefrac{1}{6}$          \\
Sr            & Sr01           & $\nicefrac{1}{2}$        & $\nicefrac{1}{2}$        & 0.18219(9) & $\nicefrac{5}{6}$          \\
Y             & Y02            & 0          & 0          & $\nicefrac{1}{2}$         & $\nicefrac{1}{6}$          \\
Sr            & Sr02           & 0          & 0          & $\nicefrac{1}{2}$         & $\nicefrac{5}{6}$         \\
Ni            & Ni03           & $\nicefrac{1}{2}$        & $\nicefrac{1}{2}$        & 0.40303(13) & $\nicefrac{7}{8}$         \\
Al            & Al03           & $\nicefrac{1}{2}$        & $\nicefrac{1}{2}$        & 0.40303(13) & $\nicefrac{1}{8}$          \\
O             & O04           & $\nicefrac{1}{2}$        & $\nicefrac{1}{2}$        & $\nicefrac{1}{2}$         & 1             \\
O             & O05           & $\nicefrac{1}{2}$        & $\nicefrac{1}{2}$        & 0.3063(7)  & 1              \\
O             & O06           & $\nicefrac{1}{2}$        & 0          & 0.4112(6)   & 0.77(4)            \\
&&&&&&\\ \hline
\textbf{Site} & \textbf{U$_{11}$}  & \textbf{U$_{22}$} & \textbf{U$_{33}$} &  \textbf{U$_{23}$} & \textbf{U$_{13}$} \\ \hline
Y01 & 0.0094(7) & 0.0094(7) & 0.0073(10) & 0 & 0\tabularnewline
Sr01 & 0.0094(7) & 0.0094(7) & 0.0073(10) & 0 & 0\tabularnewline
Y02 & 0.0237(10) & 0.0237(10) & 0.0083(13) & 0 & 0\tabularnewline
Sr02 & 0.0237(10) & 0.0237(10) & 0.0083(13) & 0 & 0\tabularnewline
Ni03 & 0.0058(9) & 0.0058(9) & 0.0086(14) & 0 & 0\tabularnewline
Al03 & 0.0058(9) & 0.0058(9) & 0.0086(14) & 0 & 0\tabularnewline
O04 & 0.015(4) & 0.015(4) & 0.010(7) & 0 & 0\tabularnewline
O05 & 0.030(6) & 0.011(5) & 0.031(6) & 0 & 0\tabularnewline
O06 & 0.017(8) & 0.017(8) & 0.011(11) & 0 & 0\tabularnewline
&&&&&&\\

\textbf{b)} &   & & &   &  \\ \hline
\textbf{Atom} & \textbf{Site}  & \textbf{x} & \textbf{y} &  \textbf{z} & \textbf{Occ} \\ \hline

Y & Y01 & $\nicefrac{1}{2}$ & $\nicefrac{1}{2}$ & 0.18212(7) & $\nicefrac{1}{6}$\tabularnewline
Sr & Sr01 & $\nicefrac{1}{2}$ & $\nicefrac{1}{2}$ & 0.18212(7) &  $\nicefrac{5}{6}$\tabularnewline
Y & Y02 & 0 & 0 & $\nicefrac{1}{2}$ &  $\nicefrac{1}{6}$\tabularnewline
Sr & Sr02 & 0 & 0 & $\nicefrac{1}{2}$ &  $\nicefrac{5}{6}$\tabularnewline
Ni & Ni03 & $\nicefrac{1}{2}$ & $\nicefrac{1}{2}$ & 0.40306(10) &  $\nicefrac{11}{12}$\tabularnewline
Al & Al03 & $\nicefrac{1}{2}$ & $\nicefrac{1}{2}$ & 0.40306(10) &  $\nicefrac{1}{12}$\tabularnewline
O & O04 & $\nicefrac{1}{2}$ & $\nicefrac{1}{2}$ & $\nicefrac{1}{2}$ & 1\tabularnewline
O & O05 & $\nicefrac{1}{2}$ & $\nicefrac{1}{2}$ & 0.3063(5) & 1\tabularnewline
O & O06 & $\nicefrac{1}{2}$ & 0 & 0.4115(5) & 0.86(3)\tabularnewline
 &  &  &  &  & \tabularnewline
 \hline
\textbf{Site} & \textbf{U$_{11}$}  & \textbf{U$_{22}$} & \textbf{U$_{33}$} &  \textbf{U$_{23}$} & \textbf{U$_{13}$} \\ \hline
Y01 & 0.0106(5) & 0.0106(5) & 0.0069(7) & 0 & 0\tabularnewline
Sr01 & 0.0106(5) & 0.0106(5) & 0.0069(7) & 0 & 0\tabularnewline
Y02 & 0.0259(9) & 0.0259(9) & 0.0077(10) & 0 & 0\tabularnewline
Sr02 & 0.0259(9) & 0.0259(9) & 0.0077(10) & 0 & 0\tabularnewline
Ni03 & 0.0087(7) & 0.0087(7) & 0.0085(10) & 0 & 0\tabularnewline
Al03 & 0.0087(7) & 0.0087(7) & 0.0085(10) & 0 & 0\tabularnewline
O04 & 0.11(2) & 0.11(2) & 0.039(17) & 0 & 0\tabularnewline
O05 & 0.018(4) & 0.018(4) & 0.006(5) & 0 & 0\tabularnewline
O06 & 0.032(7) & 0.012(5) & 0.022(6) & 0 & 0\tabularnewline
\end{tabular}

\end{table}

\begin{figure}[h]
\centering
\includegraphics[width=1.0\columnwidth]{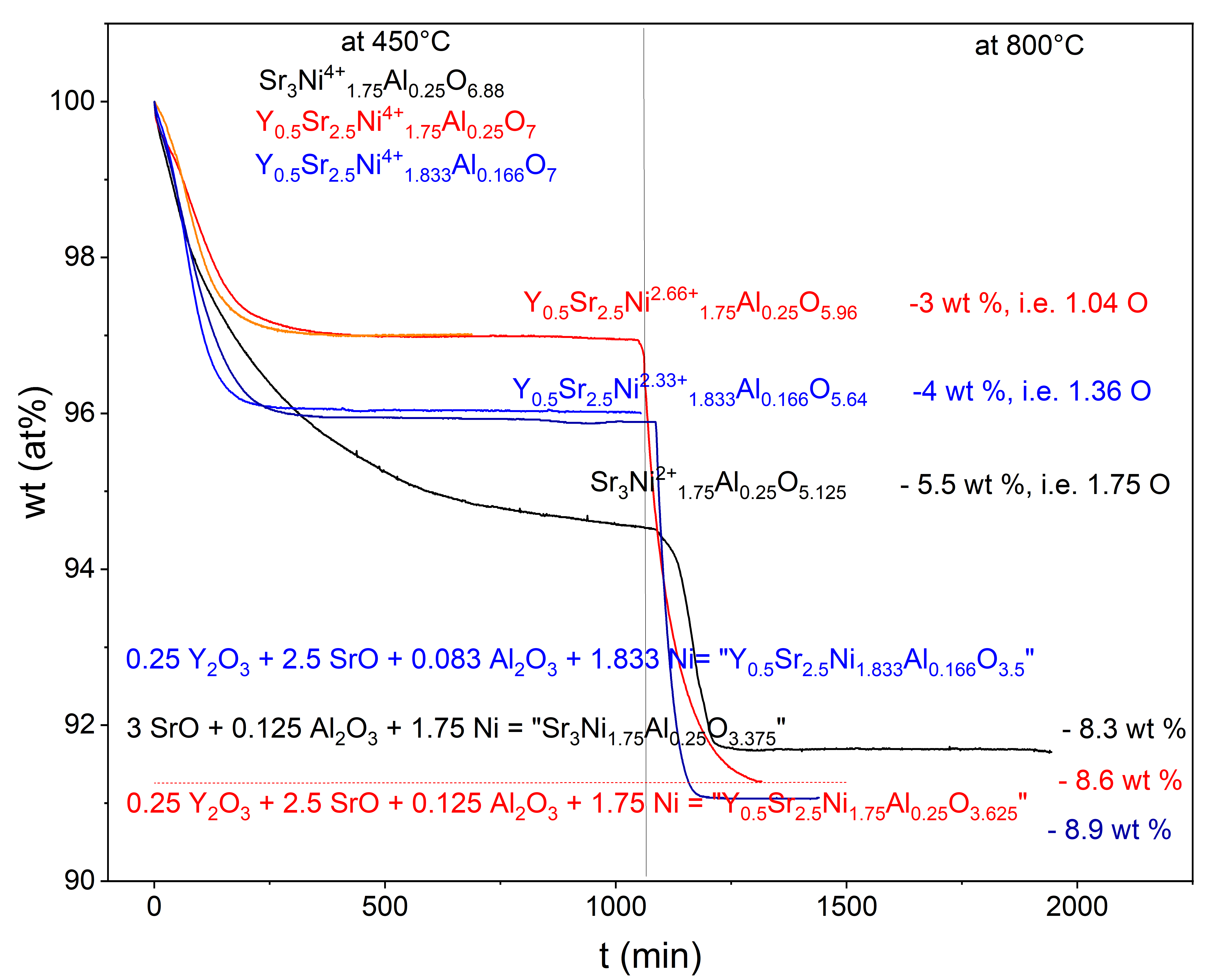}
\caption{TGA analyses of the Y$_{0.5}$Sr$_{2.5}$Ni$_{1.75}$Al$_{0.25}$O$_{7-\delta}$ (dashed and solid line show two measurements), Y$_{0.5}$Sr$_{2.5}$Ni$_{1.833}$Al$_{0.166}$O$_{7-\delta}$ (two blue lines dashed and solid) single crystals, heated at 450~\degree C (a) and 800~\degree C (b) in Ar/5\%H$_2$ flow, plotted versus time.}
\label{fig:TG_XRD}
\end{figure}

After large high quality crystals were established with $y=0.5$ and $x = 0.166$ and $x= 0.25$, their structural motif was modified by topochemical hydrogen reduction. Conventional TGA measurements were employed, and the phases were reduced in two steps, the results of which are displayed in Fig.~\ref{fig:TG_XRD}. First, the crystals were heated to 450~\degree C in 5\% hydrogen until a plateau was reached, hence a stable intermediate oxidation state. Next, the temperature was increased to 800~\degree C in order to fully decompose the species and reach metallic Ni. From the full reduction, the as-grown oxygen content was estimated via the mass loss. We can expect for the decomposed phase metallic Ni, while the other compounds remain as oxides with their stable oxidation states. Subsequently as written in Fig.~\ref{fig:TG_XRD} we reduce down to an oxygen stoichiometry of 3.625 and 3.5 for YSNAO $x = 0.166$ and $x= 0.25$, respectively. This corresponds to a molar mass of 430.96 g/mol and 431.58 g/mol. We observe a weight loss of  8.83~wt\% and 8.9~wt\%. Hence the beginning molecular weight can be calculated by  The intermediate oxidation state can be calculated by 430.96 g/mol $\cdot 100/(100-8.83) = 472.7$ g/mol and 473.74 g/mol. From that we have to subtract the molecular weight of the elements Y, Sr, Ni and Al amounting to 372.96 g/mol and 375.58 g/mol. Hence we have a total oxygen molecular weight of 99.74 g/mol and 98.16 g/mol. Resulting in a stoichiometry of 6.23 and 6.13 with a nominal oxidation state of Ni$^{2.7+}$ and Ni$^{2.875+}$. These estimations and the corresponding Ni oxidation state are summarized in Table~\ref{tab:lattice_comp} and supported by the oxygen EDX results as well as the oxygen refinement from single crystal XRD summarized in Table~\ref{tab:SNOatom_pos}.

For all phases we find a metastable topochemical phase with an intermediate oxidation state realized, with individual weight losses. Due to the different amount of dopings varying oxygen losses are seen with 5.83~wt\%, and 4.9~wt\%. Leading to a reduced stoichiometry of YSNA$_{0.25}$O$_{5.29}$ and YSNA$_{0.166}$O$_{4.89}$. For all reduced phases, the same orthorhombic crystal structure of $Immm$ was observed both via powder refinement and single crystal refinement (as shown later), where the in-plane oxygen removal is seen in the reduction of the lattice constant $a$ combined by a subtle increase of $b$ and a large increase of the lattice constant $c$ (see Table~\ref{tab:lattice_comp}).

\begin{figure}[tb]
\includegraphics[width=1.0\columnwidth]{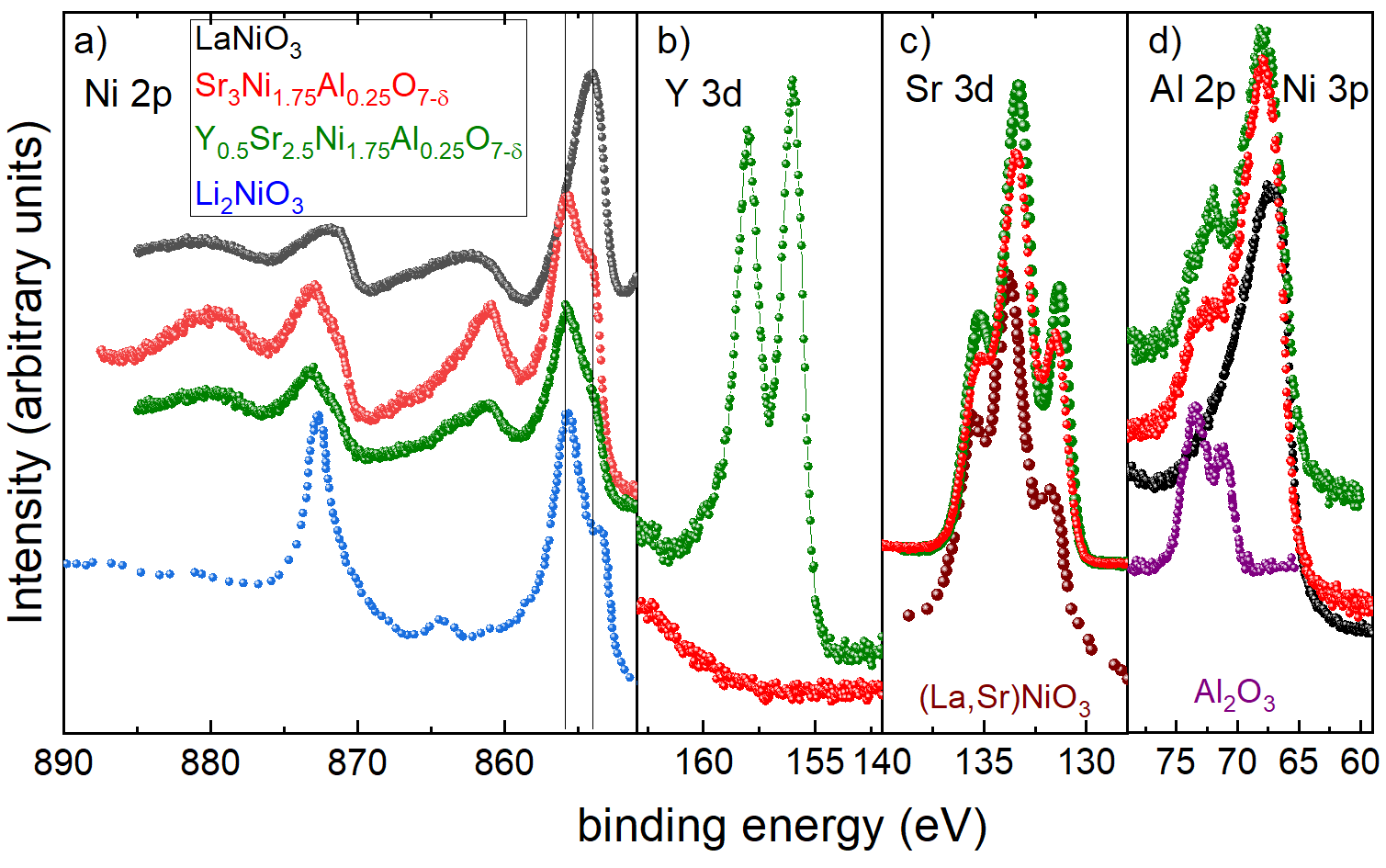}
\caption{XPS spectra of LaNiO$_{3}$ \cite{Puphal2023APL} (black), SNAO \cite{Yilmaz2024} (red) and Y$_{0.5}$Sr$_{2.5}$Ni$_{1.75}$Al$_{0.25}$O$_{7-\delta}$ (green) for mutual comparison. These are displayed in the energy ranges corresponding to the binding energies of (a) Ni $2p$, (b) Y $3d$, (c) Sr $3d$ and (d) Al~$2p$ states. The XAS spectrum of Li$_2$NiO$_3$ \cite{Bianchini2020} is shown for reference.
}
\label{XPS}
\end{figure}

To confirm oxidation state in the doped system, XPS analyses were carried out. The corresponding results are displayed in Fig.~\ref{XPS}, where spectra within the relevant binding energies are shown for (a)~Ni $2p$, (b)~Y $3d$, (c)~Sr $3d$, and (d)~Al $2p$. In the Ni case for YSNAO, the strongest peak appears at 855.6~eV. A shoulder at 854.0~eV is visible as well, where both are indicated by a vertical line. Here, the shoulder is more intense compared to the one in the reference  $\mathrm{Li}_2\mathrm{NiO}_3$ XAS spectrum \cite{Bianchini2020}. For LaNiO$_3$ there are majorly overlapping La lines and the two Ni lines have also been interpreted in literature as $\mathrm{Ni}^{2+}$ and $\mathrm{Ni}^{3+}$ \cite{Hu2023}. Similarly, $\mathrm{Sr}_{2.5}\mathrm{Y}_{0.5}\mathrm{Ni}_{1.85}\mathrm{Al}_{0.15}\mathrm{O}_{7-\delta}$, the Ni $2p$ spectra has the same character as for SNAO, but a slightly weaker feature is associated with  the peak at ~854~eV. 
Notably, the reference $\mathrm{Li}_2\mathrm{NiO}_3$ and the XPS data differed by the presence of an additional satellite around 861~eV, which was attributed to a residual amount of decomposing hydroxides on the examined samples' surfaces. 
The evaluation of Ni oxidation states cannot be simply attributed to a single line and its evolution from 1+ to 4+ is rather complex. A recent study investigated LaNiO$_2$, NiO, LiNiO$_2$ and Li$_2$NiO$_3$ via XAS \cite{Fajardo2025}. For  oxidation states 2-4+ Ni reveals these two main lines, where the intensity of the 854 eV line gradually decreases and the intensity of the 855.6~eV increases with oxidation state increase among a visible charge transfer \cite{Fajardo2025}. 
In Fig.~\ref{XPS}~(b), for Y $3d_{5/2}$ and $3d_{3/2}$, the expected two spin-orbit split peaks were observed at 156 and 158~eV, respectively, corresponding to its 3+ oxidation state \cite{Kirikova2002}. For the Sr $3d$ case shown in Fig.~\ref{XPS}~(c), two doublets appear: first at the Sr $3d_{5/2}$ binding energy, at 131.5~eV, then at 133.5~eV. 
The latter can be related to the instability of Sr on the surface, probably reacting to form Sr-oxide or hydroxide \cite{Wang2018}. The spectra is very similar to that of perovskite (La,Sr)NiO$_3$ single crystals \cite{Puphal2023APL}. Notably, the Al $2p$ lines (with $\mathrm{Al}_2\mathrm{O}_3$ as a reference \cite{TAGO2017}) are slightly better resolved for YSNAO sample, as visible in Fig.~\ref{XPS}~(d), but their overlap with Ni $3p$ features makes a clear analysis impossible.
A relative change of intensity to the higher energy for Ni 2p indicates a high oxidation state. However, the XPS yields no definite statement for the Ni oxidation state, but a clear deviation from simple Ni$^{2+}$ is visible.
Consequently, we have also used the determination of oxygen via the EDX spectra (which were determined on two device to allow for better reliability) to estimate the Ni oxidation state as shown in Table~\ref{tab:lattice_comp}. In addition we have refined the oxygen occupancy in our single crystal refinement, which is also summarized in Table~\ref{tab:SNOatom_pos}. Notably, for SNAO we have conducted a gas extraction measurement \cite{Yilmaz2024} yielding an oxygen content of 6.88(5) yielding a higher and more trustful result than EDX which for the light element of oxygen is not really reliable. Similarly, oxygen occupancy refinement is extremely sensitive and might rather yield too low contents. TGA is quite reliable but even here, we assume a full decomposition. While post XRD reveals Ni, SrO is instable and was not detected in XRD leaving an uncertainty of the exact decomposition. Here, any deviation would hint at a higher oxidation state than determined, as the expected stoichiometry is the lowest the system can reach. Combining these data we expect an oxidation state around 3+ for YSNAO.

\subsection{Effectivity of Doping}

\begin{figure}[tb]
\includegraphics[width=1.0\columnwidth]{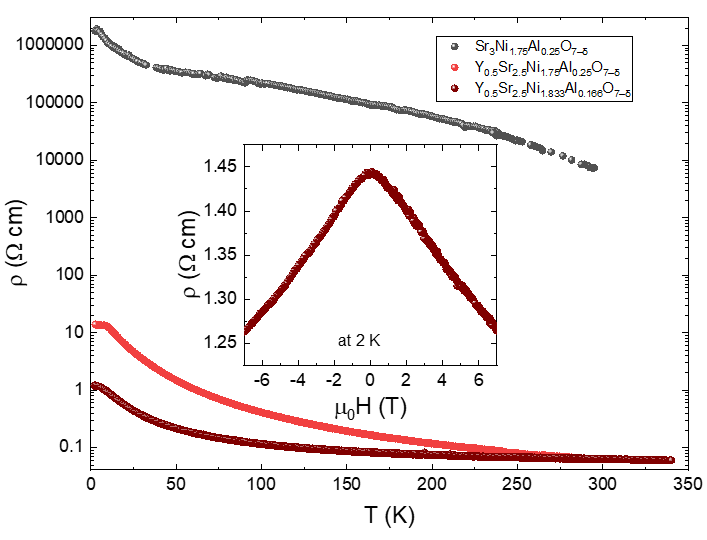}
\caption{Electrical transport for YSNAO. Shown are the resistivity of Sr$_{3}$Ni$_{1.75}$Al$_{0.25}$O$_{7-\delta}$ (black), Y$_{0.5}$Sr$_{2.5}$Ni$_{1.75}$Al$_{0.25}$O$_{7-\delta}$ (red) and Y$_{0.5}$Sr$_{2.5}$Ni$_{1.833}$Al$_{0.166}$O$_{7-\delta}$ (brown) versus temperature in the range between 1.8 and 300~K. As an inset, the magnetoresistance of Y$_{0.5}$Sr$_{2.5}$Ni$_{1.833}$Al$_{0.166}$O$_{7-\delta}$ is shown under fields varying from -7 to 7 T, at a temperature of 2~K.}
\label{R}
\end{figure}

Finally, the electrical properties of the novel RP-type nickelate crystals were measured. While our DFT calculations predict a metallic ground state, we found that the introduction of Al, which enable the stabilization of the phase, reduces the conductivity tremendously \cite{Yilmaz2024}. This likely occurs as Al usually has a tetrahedral oxygen environment, while the one given in the structure is an octahedral one. Thus, the large amount of oxygen disorder found in refinement could explain this rise in resistivity. Fig.~\ref{R} presents the resistivity of three crystals, where SNAO exhibits insulating-semiconducting behavior with a wide band gap. With the introduction of $y=0.5$ amount of Y to the compound, a drastic reduction in the resistivity by six orders of magnitude was observed. Similarly, when reducing the amount of Al from $x=0.25$ to $x=0.166$, the resistivity at room temperature was nearly unchanged but reduced at low temperatures. Nonetheless, even for the lowest Al content $x=0.16$, an increasing resistivity with decreasing temperature was still observed. A magnetic transition remains present in all compounds at low temperatures, appearing  a weak downturn in resistivity. A very small magnetoresistance was observed under an applied field at 2~K, as shown in the inset of Fig.~\ref{R}. In order to estimate the band gap for these systems, the Arrhenius approach was used by plotting the resistivity $\rho$ as $\ln{\rho}$ against a temperature $T$ axis as $1/T$, fitting the low-temperature part as for SNAO \cite{Yilmaz2024}. Here, all carriers are thermally excited across the band gap, and the electron and hole concentrations are equal. The fit using $\rho = \rho_{0} e^{E_{g}/kT}$ yields 1250(30)~K, 394(2)~K and 103(1)~K for $\mathrm{Sr}_{3}\mathrm{Ni}_{1.75}\mathrm{Al}_{0.25}\mathrm{O}_{7-\delta}$, $\mathrm{Y}_{0.5}\mathrm{Sr}_{2.5}\mathrm{Ni}_{1.75}\mathrm{Al}_{0.25}\mathrm{O}_{7-\delta}$ and $\mathrm{Y}_{0.5}\mathrm{Sr}_{2.5}\mathrm{Ni}_{1.833}\mathrm{Al}_{0.166}\mathrm{O}_{7-\delta}$, respectively.

\begin{figure}[tb]
\includegraphics[width=1.0\columnwidth]{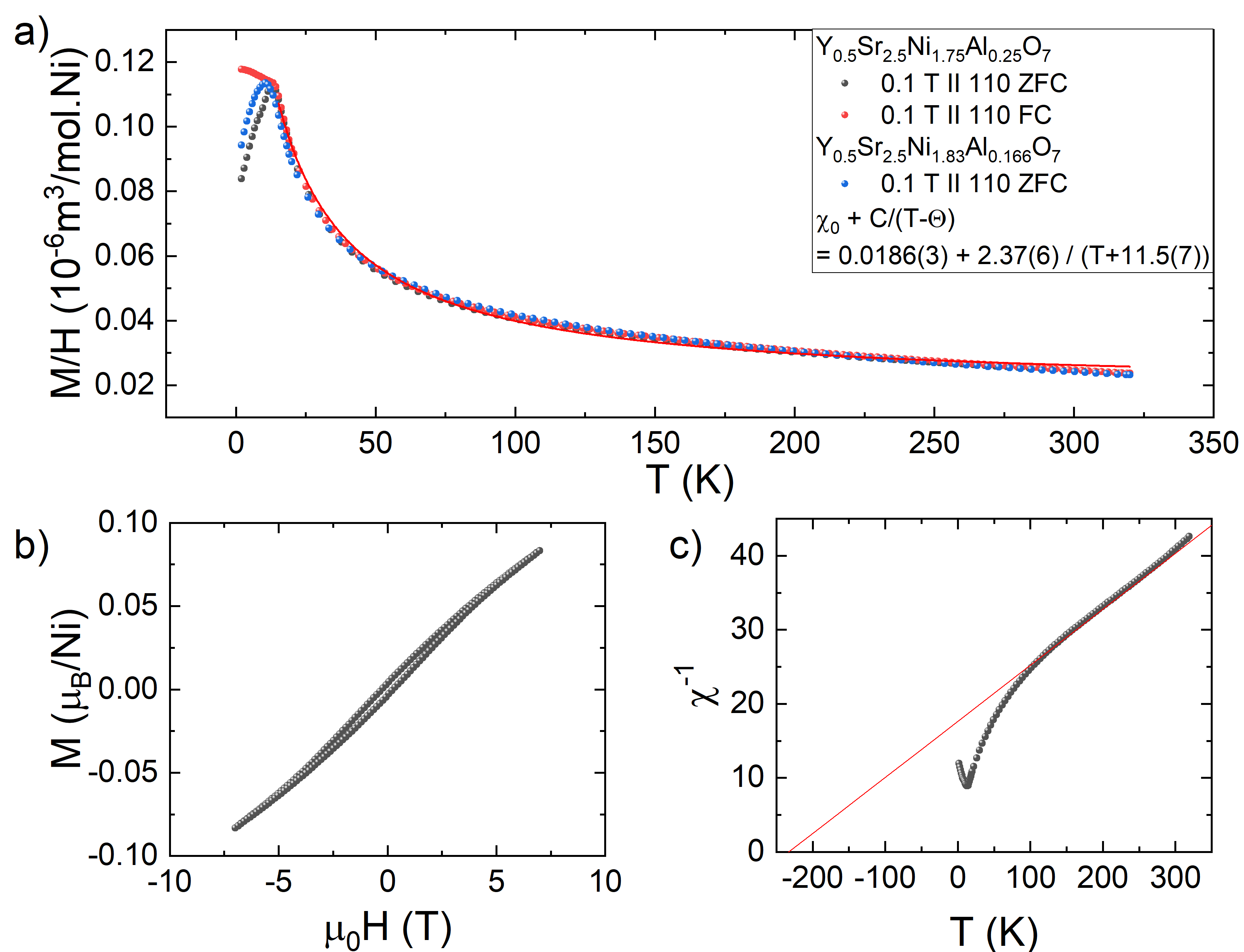}
\caption{Magnetic properties of a Y$_{0.5}$Sr$_{2.5}$Ni$_{1.75}$Al$_{0.25}$O$_{7}$ crystal and a Y$_{0.5}$Sr$_{2.5}$Ni$_{1.833}$Al$_{0.166}$O$_{7}$ crystal: (a) Magnetization versus temperature at an external field of 0.1 T applied along 110 (black) and the $c$-axis (red). (b) Magnetization versus field at 2 K for field along 110 (black) and the $c-$axis (red). (c) Inverse magnetization versus temperature of the data in c) with Curie-Weiss fits as lines.
}
\label{M}
\end{figure}

\begin{figure}[h]
\centering
\includegraphics[width=1.0\columnwidth]{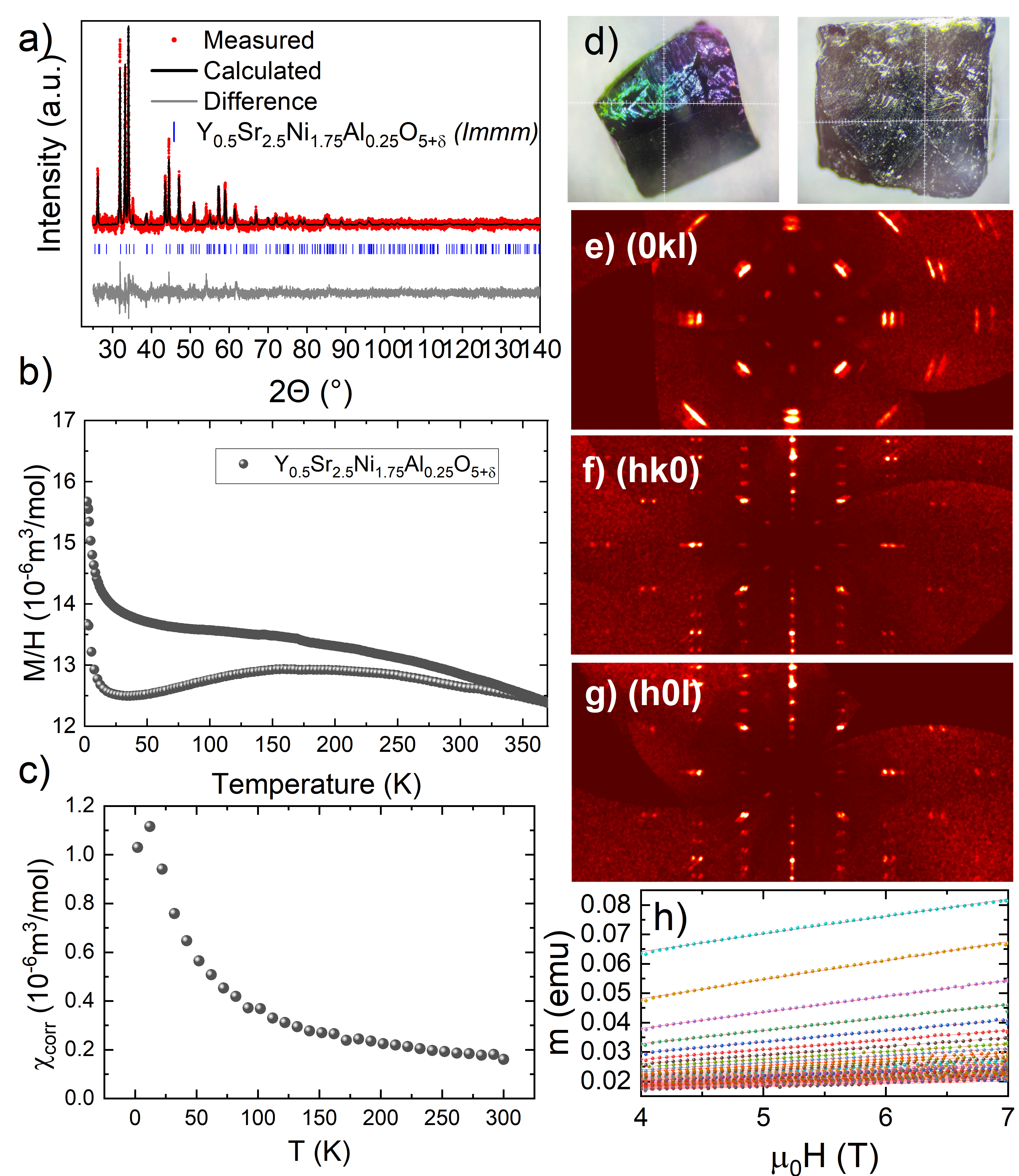}
\caption{Reduced crystals. (a) Powder diffraction pattern of a crushed reduced YSNAO single crystals. Rietveld fit of the room-temperature PXRD pattern of $\mathrm{Y}_{0.5}\mathrm{Sr}_{2.5}\mathrm{Ni}_{1.75}\mathrm{Al}_{0.25}\mathrm{O}_{5+\delta}$. The solid black line corresponds to the calculated intensity from the Rietveld refinement, the solid gray line is the difference between the experimental and calculated intensities, and the vertical blue/green bars are the calculated Bragg peak positions. (b) Magnetization versus temperature at an external field of 0.1 T for a reduced YSNAO and SNAO crystal. (c) Magnetization versus temperature in an external field of 7 T for YSNAO. (d) Crystal image and its single crystal XRD zonal diffraction maps (e) (hk0), (f) (0kl) and (g) (h0l) integrated with the $Immm$ cell.
}
\label{reduced}
\end{figure}

The crystals cleave well giving rectangle pieces, where the $c-$axis lies out of the plane, and the long direction is the [110] direction with dimensions of 7 x 3 x 3~mm$^3$. Magnetic susceptibility measurements are displayed in Fig.~\ref{M} for the compound with phase pure single crystals of $x=0.166$ and $x=0.25$. Over all Al-dopings, the crystals reveal an antiferromagnetic ordering with a transition around 13~K and 11~K, and 15~K for SNAO \cite{Yilmaz2024}. Since AC-susceptibility showed no frequency dependence and no signal in $\chi''$, antiferromagnetic ordering was concluded, as previously established for SNAO \cite{Yilmaz2024}. Notably, with the increase in the size of the obtained single crystals, the Curie-Weiss fits can be improved with reliable masses and signals. In Fig.~\ref{M}~(c) for the $y = 0.5$ and $x=0.25$ crystal, a Curie-fit with the standard linear fitting procedure for the inverse susceptibility in a range of 100-300 K is shown. This fit gives a Weiss temperatures of -240~K found along the [110] direction. The fit yields a decreased slope of 0.07568(2), resulting in a magnetic moment  $\mu_{\mathrm{eff}}=2.89\,\mu_\mathrm{B}$. If we instead apply a Curie-fit in the whole temperature range of 15-320~K of $\chi_0+\frac{C}{T-\Theta}$ we find a reduced Weiss temperature of 11.5(7)~K. The Curie constant is reasonable with 2.37(6)~cm$^3$~K~/mol. Ni, however the $\chi_0$ is unreasonably large with 1.5x10$^{-3}$~emu/Oe, possibly originating from ferromagnetic impurities. This fit would suggest a simple magnetic transition at the Weiss temperature of around 11.5(7)~K from a S=1 system, with dominant Ni$^{2+}$.
After reduction the crystals undergo a structural transition towards the orthorhombic $Immm$ structure solved both by powder XRD and single crystal XRD shown in Fig.~\ref{reduced}~(a), (e),(f) and (g). Including Ni as a secondary phase we find 3.3(3)~wt\%, with all peaks overlapping with the $Immm-$phase. The crystals reduce to twinned domains with shared $c-$axis, which still reveal high crystallinity. This is visible in the fourier maps and the crystal pictures shown in Fig.~\ref{reduced}~(d). However their properties are completely dominated by ferromagnetic Ni impurities below the detection limit of XRD and spin glassiness from enormous oxygen disorder similar to reduced perovskite phases \cite{Puphal2021,Ortiz2022,Puphal2023B}. By a strong bifurcation of the field cooled and zero field cooled curve shown in Fig.~\ref{reduced}~(b) already occuring above 370~K. In the following, we address this issue by susceptibility measurements in strong external fields. In the case of ferromagnetic impurities in reduced samples, it is expected that their magnetization signal saturates in sufficiently strong fields, yielding only a constant contribution after the field exceeds a certain threshold
value. Moreover, large fields suppress the glassy dynamics and underlying intrinsic magnetic correlations can be exposed. Magnetization-field isotherms $M(H)$ were measured at various temperatures (Fig. \ref{reduced} h). The saturation
magnetization $M_ {sat}$ can be determined from linear fits to the isotherms at high fields in the range between 4 and 7 T. Via the Honda-Owen
method, which extrapolates the measured susceptibility M/H = $\chi_{corr} + C_{sat}M_{sat}/H$ for $1/H \rightarrow 0$, where $M/H$ is the measured susceptibility, $C_{sat}$ the presumed ferromagnetic impurity content, and
$M_{sat}$ its saturation magnetization, the  corrected, intrinsic susceptibility $\chi_{corr}$ can be extracted. In Fig. \ref{reduced} c) the temperature dependence of $\chi_{corr}$, which is
drastically different from $M/H$ in Fig. \ref{reduced} b), is shown. After correction the susceptibility again reveals the magnetic transition around 13 K as for the non-reduced compound, with a slightly increased signal from the increased amount of Ni${2+}$ suggesting magnetic order along the Ni-O-Ni chains.

Notably, ligand-hole localization is common in transition-metal oxides with unusually high valence states; in such cases, when temperature decreases, the ligand holes lose kinetic energy and localize. Future studies of low-temperature neutron diffraction on these now large high-quality crystals will aim to explain the observed low magnetic moment.

\section{Summary}
In conclusion, OFZ growth can be readily used to investigate phase diagrams even for high gas pressures. A novel doping-tuned synthesis of YSNAO crystals was demonstrated, reducing the band gap, where the conductivity increased in the order of $10^6$ for Y contents of $y = 0.5$ and Al $x = 0.166$. In all cases the system shows an antiferromagnetic order below 15~K and the magnetization indicates a low moment corresponding to a $S=1$ system. TGA measurements show a wide range of Ni oxidation states dependent on the doping and oxygen content in the system, similarly to the case of rare-earth RP-type nickelate superconductors with varying $n$. The high crystallinity and quality of these crystals with neither polymorphism nor tetragonal-to-orthorhombic transitions –characteristic of rare-earth-based systems– make them exceptional. Notably, to explore superconductivity the lowest necessary co-substitution of 8.75\% Al on the Ni-site still has to be overcome, but our DFT promise an exciting family of compounds for ambient pressure superconductivity. The large crystals allow for future neutron studies, which might reveal a ligand hole formation known from various high oxidation state transition metal systems possibly explaining the low moment.

\section*{acknowledgments}
We thank the Solid State Spectroscopy department for the use of their SQUID and PPMS systems.
\section*{AUTHOR DECLARATIONS}
\subsection*{Conflict of Interest}
The authors have no conflicts to disclose.
\subsection*{Author Contributions}
P.P. conceived the project, supervised the experiments and measured transport and magnetization. H.Y. grew the crystals, performed XRD, EDX and TGA measurements under the supervision of P.P. P.S.L. prepared the STEM lamella, performed the STEM measurements under the supervision of Y.E.S. The authors M.I., O.C. and P.v.A. were responsible for the project management. DFT calculations were performed by M.K. XPS samples were prepared by P.P., measured and analyzed by K.K. and U.S. The authors H.Y. and P.P. wrote the manuscript with comments from all authors.
\subsection*{DATA AVAILABILITY}
The data that support the findings of this study are available from the corresponding author upon reasonable request.

\bibliography{Literature}

\end{document}